\documentclass[11pt]{article}
\usepackage{amsmath,amssymb}

\def\whitebox{{\hbox{\hskip 1pt
        \vrule height 6pt depth 1.5pt
        \lower 1.5pt\vbox to 7.5pt{\hrule width
                  3.2pt\vfill\hrule width 3.2pt}%
        \vrule height 6pt depth 1.5pt
        \hskip 1pt } }}
\def\qed{\ifhmode\allowbreak\else\nobreak\fi\hfill\quad\nobreak\whitebox\medbreak}

\newtheorem{theorem}{Theorem}
\newtheorem{corollary}{Corollary}
\newtheorem{lemma}{Lemma}
\newtheorem{proposition}{Proposition}

\newtheorem{definition}{Definition}
\newtheorem{remark}{Remark}
\newtheorem{construction}{Construction}
\newtheorem{example}{Example}

\begin{document}
\title{Secondary Constructions of Bent Functions and Highly Nonlinear Resilient Functions}
\author{Fengrong~Zhang \footnote{School of Computer Science and Technology,
China University of
 Mining and Technology, Xuzhou, Jiangsu 221116, P.R.
China (e-mail:
zhfl203@cumt.edu.cn)}~ ~Claude~Carlet \footnote{Department of Mathematics, LAGA, UMR 7539, CNRS,  Universities of Paris 8 and Paris 13, 93526 Saint-Denis cedex 02, France (claude.carlet@inria.fr)}~ ~Yupu~Hu \footnote{State Key Laboratory of Integrated
Services Networks, Xidian University,
Taibai Road 2, Xi'an 710071, P.R. China (e-mail:
yphu@mail.xidian.edu.cn)}~~ Wenzheng~Zhang \footnote{Science and Technology on Communication Security Laboratory,
the 30th Research Institute of China Electronics Technology Group Corporation,
   Chengdu 610041, P.R. China (e-mail:
zwz85169038@sina.com)}}



%
%

\markboth{}
{Shell \MakeLowercase{\textit{et al.}}: Bare Demo of IEEEtran.cls
for Journals}
%



\maketitle

\begin{abstract}
  In this paper, we first present a new secondary construction of bent functions (building new bent functions from
two already defined ones).  Furthermore, we apply the  construction using as  initial functions some specific bent functions and then  provide several concrete constructions of bent functions.
   The  second part of the paper is devoted to the constructions of resilient functions.   We give a generalization of the indirect sum construction for constructing resilient functions with high nonlinearity.  In addition, we modify the generalized construction to ensure a high nonlinearity of the constructed function.
\end{abstract}

{\bf Keywords : }
  Boolean function,
   bent function, resilient function, high nonlinearity.

%

\section{Introduction}
%
%
%
%
Bent  functions were introduced
  by Rothaus in 1976 as an interesting combinatorial object with the
  important property of having optimal nonlinearity \cite{O.S.Rothaus}.
  Since bent functions  have many applications in sequence design, cryptography and algebraic coding
  \cite{F.J. MacWilliams,J. D. Olsen},
   they have been extensively studied
   during the thirty last years
  \cite{C. Carlet1994,CC-C-GPS,CC-CDL,J. Dillon,CC-Dobbertin-bis,CC-guillot,Q. Meng,CC-Wolf99}.
   In terms of  sequence design, several binary bent sequences were constructed by using the bent functions \cite{J. S. No,J. D. Olsen}.  Binary bent sequences can be good candidates for many commutation systems such as code-division multiple-access systems, radar systems, and synchronization systems in that they  have optimal correlation and balance property \cite{A. Lempel,J. S. No,J. D. Olsen}. In addition,  bent functions can also be
   used to construct highly nonlinear balanced functions  \cite{H. Dobbertin}.

With regard to constructions of bent functions, there are two kinds of constructions: primary constructions (designing functions without using
known ones) and secondary constructions.
 The primary constructions mainly include the Maiorana-McFarland (M-M) class
\cite{J. Dillon},  the partial spreads (PS) class \cite{J.
Dillon} and  Dobbertin gave a construction of a
class of bent functions which leads to some elements of M-M class
and of PS class as extremal cases \cite{H. Dobbertin}.
The secondary constructions mainly include direct sum construction \cite{J. Dillon}, Rothaus' construction \cite{O.S.Rothaus}, indirect sum construction \cite{C.Carlet2004-bis}.
Moreover,
there are some constructions of bent functions proposed in
\cite{C. Carlet1994,Carlet1996,C. Carlet2004,Carlet2012Zhang,G.
Leander}.
However, although many concrete constructions of bent
functions have been discovered,
 the general structure of bent
functions is still unclear. In particular a complete classification
of bent functions seems hopeless today.

 Resilient functions have important applications in the nonlinear combiner model of stream cipher \cite{Camion92,Siegenthaler84,Xiao88}. Over the last decades,  much attention was paid to the construction
of highly nonlinear Boolean functions
in the cryptographic literature
\cite{Carlet2002,Fu2012,Pasalic2006,Sarkar2000,Zeng2005,Zhang2009,Zhang2012SCN,Zhang2012CM}.
   In terms of constructions of resilient functions, there are also  two kinds of constructions which are primary constructions and secondary constructions.
 The primary constructions mainly include
Maiorana-McFarland's construction \cite{Camion92}, generalizations of Maiorana-McFarland's construction \cite{Carlet2002,Carlet2005 DCC}, Dobbertin's construction \cite{H. Dobbertin,Seberry1994} and other constructions \cite{Fu2011,Zhang2009}. In addition, the simple secondary constructions mainly include direct sum of functions \cite{Siegenthaler84},  Siegenthaler's construction \cite{Siegenthaler84}, Tarannikov's elementary construction \cite{Tarannikov2000}, indirect sum of functions \cite{C.Carlet2004-bis} and constructions without extension of the number of variables \cite{Carlet AAECC}.
Many highly nonlinear Boolean functions can be constructed by using the above constructions.

  In this paper, we first present a new secondary construction of bent functions.  We show how to construct an $(n+m-2)$-variable bent function from two known bent functions in $n$ variables and
   in $m$ variables respectively.  Furthermore, by selecting the known bent functions as  the initial functions of the new secondary construction, we can provide several concrete constructions of bent functions which include primary constructions (Corollary \ref{construction bent nMM} and Corollary \ref{construction bent D})  and secondary constructions (Corollary \ref{construction bent by PSab} and Corollary \ref{construction bent Rothaus}).
   In the second part of the paper, we present a generalization of the indirect sum construction for constructing resilient functions with high nonlinearity.   On this basis,  we provide another two   secondary constructions of resilient functions.  It is  shown that  many new $(n+m)$-variable functions with nonlinearity strictly more than $2^{n+m-1}-2^{\lfloor(n+m)/2\rfloor}$ can be easily obtained by using these secondary constructions, where $\lfloor(n+m)/2\rfloor$ denotes the largest integer not exceeding $(n+m)/2$.

The rest of the paper is organized as follows. Section \ref{2}
introduces basic definitions and cryptographic criteria relevant for
Boolean functions. In Section  \ref{3}, we present a method for constructing
bent
 functions. In Section \ref{4}, we provide a generalization of the indirect sum construction for constructing resilient functions.
 At last, some conclusions are given in Section
 \ref{5}.

 \section{Preliminaries}\label{2}

In the remainder of this paper,  we denote the additions and multiple sums
over the finite field ${\Bbb F}_2$  by $\oplus$ and $\bigoplus$.
 Let ${\Bbb F}_{2}^{n}$ be the
$n$-dimensional vector space over ${\Bbb F}_{2}$, and  $B_{n}$  the
set of all $n$-variable Boolean functions from ${\Bbb F}_{2}^{n}$ to
${\Bbb F}_{2}$.
  A basic representation of a Boolean function
$f(x_{1},\ldots,x_{n})$   is by the output column of its
truth-table, i.e., a binary string of   length $2^{n}$,
\begin{displaymath} \begin{array}{c}
  [f(0,\ldots,0,0,0),
 \ldots, f(1,\ldots,1,1,0), f(1,\ldots,1,1,1)].
   \end{array}
   \end{displaymath}

     The \emph{Hamming weight} wt($f$) of a Boolean function $f\in
  B_{n}$ is the weight of the above binary string. We say a Boolean function $f$
  is \emph{balanced} if its Hamming weight equals $2^{n-1}$. The
 \emph{ Hamming distance} $d(f,g)$ between two Boolean functions $f$
 and $g$ is the Hamming weight of their difference $f\oplus g$.

    Any Boolean function has a unique representation as a
 multivariate polynomial over ${\Bbb F}_{2}$, called the \emph{algebraic normal
 form}(ANF):
 \begin{displaymath}
  f(x_{1},\ldots,x_{n})=\bigoplus_{I\subseteq\{1,2,\ldots,n\}}a_{I}\prod_{l\in I}x_{l}
 \end{displaymath}
 where  $a_{I}\in {\Bbb F}_{2}$, and the terms $\prod_{l\in I}x_{l}$
  are called monomials. The \emph{algebraic degree} $\deg(f)$ of a
  Boolean function $f$ equals the maximum degree of those monomials
  whose coefficients are nonzero in its ANF. A
  Boolean function is affine if it has algebraic degree at most $1$. The set of
 all $n$-variable affine functions is denoted by $A_{n}$. An $n$-variable affine
 function with
  constant term $0$ is a linear function, and can be represented as
   $\omega \cdot x=\omega_{1}x_{1}\oplus\ldots\oplus\omega_{n}x_{n}$
  where $\omega=(\omega_{1},\ldots,\omega_{n})\in {\Bbb F}_{2}^{n},
  x=(x_{1},\ldots,x_{n})\in {\Bbb F}_{2}^{n}$.

  The \emph{nonlinearity} of $f\in B_{n}$ is its distance
  to the set of all $n$-variable affine functions, i.e.,
\begin{displaymath}
\begin{array}{c}
      N_{f}=\min\limits_{g\in A_{n}}d(f,g).
\end{array}
  \end{displaymath}
  Boolean functions used in cryptographic systems must have high
  nonlinearity to withstand linear and fast correlation attacks  \cite{Canteaut2000}.

The \emph{Walsh transform} of
  $f\in B_{n}$ is the integer valued function over $ {\Bbb F}_{2}^{n}$ defined as
  \begin{displaymath}
      W_{f}(\omega)=\sum_{x\in {\Bbb F}_{2}^{n}}(-1)^{f(x)\oplus \omega\cdot x}.
  \end{displaymath}
   In terms of Walsh spectrum, the nonlinearity of $f$ is given by
  \begin{displaymath}
      N_{f}=2^{n-1}-\frac{1}{2}\max_{\omega\in
      {\Bbb F}_{2}^{n}}|W_{f}(\omega)|.
  \end{displaymath}


    Parseval's equation \cite{F.J. MacWilliams} states that
$ \sum_{\omega\in {\Bbb F}_{2}^{n}}(W_{f}(\omega))^{2}=2^{2n}$
 and implies that
$$ N_f \leq 2^{n-1}-2^{n/2-1}.$$

\begin{definition} \label{bent def}\cite{J. Dillon,O.S.Rothaus}
A Boolean function $f\in B_n$ is called bent  if $W_f(a)=\pm
2^{n/2}$ (that is, $ N_f = 2^{n-1}-2^{n/2-1}$) for every $ a\in \Bbb{F}_2^{n}$ ($n$ even).
\end{definition}

 If $f\in B_n$ is bent, then the dual function $\widetilde{f}$ of
 $f$, defined on ${\Bbb F}_2^n$ by:
 \[W_f(\omega)=2^{n/2}(-1)^{\widetilde{f}(\omega)}\]
 is also bent and its own dual is $f$ itself.

\begin{definition}\label{def plateaued} \cite{Zheng1999}
Let $f\in B_n$. If there exists an even integer $r$, $0\leq r\leq
n$, such that $\parallel\{\omega|W_{f}(\omega)\neq 0, \omega\in
{\Bbb F}_2^n\}\parallel=2^r$, where $\parallel \cdot\parallel$ denotes the size of a set, and $(W_{f}(\omega))^2$ equals $2^{2n-r}$ or $0$, for every $\omega\in {\Bbb F}_2^n$,
then $f$ is called an $r$th-order plateaued function in $n$
variables.  If
$f$ is a $2\lceil \frac{n-2}{2}\rceil$th-order plateaued function in $n$
variables, where $\lceil n/2\rceil$ denotes the smallest integer
exceeding $n/2$, then $f$ is also called a semi-bent function.
\end{definition}

   A Boolean function $f\in B_{n}$ is said to be \emph{correlation-immune
   of order $r$ $(1\leq r\leq n)$},
   if the output of $f$ and any $r$ input variables are statistically
   independent. Balanced
  $r$th-order correlation
  immune functions are called \emph{$r$-resilient} functions. The set of $r$th-order correlation
  immune (resp. $r$-resilient) Boolean functions is included in that of $(r-1)$th-order correlation
  immune (resp. $(r-1)$-resilient) Boolean functions.
   The correlation immunity (resp. resiliency) can also be characterized  by using the
   Walsh transform domain \cite{Xiao88}:
   \begin{lemma}\label{xiao messay}
   Let $f\in B_n$, then  $f$ is $r$th-order correlation
  immune (resp. $r$-resilient)
  if and only if its Walsh transform satisfies
  $W_{f}(\omega)=0,$ for all $\omega\in F_2^n$ such that
   $1\leq $\emph{wt}$(\omega)\leq r$ (resp. $0\leq $\emph{wt}$(\omega)\leq r$).
  \end{lemma}
  \emph{Siegenthaler's Inequality} \cite{Siegenthaler84} states that any $r$th-order correlation
       immune function has degree at most $n-r$, that $r$-resilient
       function ($0\leq r \leq n-1$) has degree smaller than or
       equal  $n-r-1$ and that any $(n-1)$-resilient function has
       algebraic degree 1.   Sarkar and Maitra \cite{Sarkar2000} have shown that the nonlinearity of any $m$-resilient function ($m \leq n-2$) is divisible by
$2^{m+1}$ and is therefore upper bounded by $2^{n-1}-2^{m+1}$. If a function achieves this
bound (independently obtained by Tarannikov \cite{Tarannikov2000} and Zheng and Zhang \cite{zheng2000}),
then it also achieves Siegenthaler's bound (cf. \cite{Tarannikov2000}).  More precisely, if
$f$ is $m$-resilient and has algebraic degree $d$, then its nonlinearity is divisible by $2^{m+1+\lfloor\frac{n-m-2}{d}\rfloor}$ (see \cite{Carlet2001,CC-Sarkar}) and can therefore be equal to $2^{n-1}-2^{m+1}$ only if $d=n-m-1$. Moreover, if an $m$-resilient function achieves nonlinearity $2^{n-1}-2^{m+1}$, then
the Walsh spectrum of
the function has then three values (such functions are often called ``plateaued''
or ``three-valued'').  We shall say that an $m$-resilient function
achieves the best possible nonlinearity if its nonlinearity equals $2^{n-1}-2^{m+1}$.
If $2^{n-1}-2^{m+1}$ is greater than the best possible nonlinearity of all balanced
functions (and in particular if it is greater than the best possible nonlinearity
$2^{n-1}-2^{n/2-1}$ of all Boolean functions) then, obviously, a better bound
exists. In the case $n$ is even, the best possible nonlinearity of all balanced
functions being smaller than $2^{n-1}-2^{n/2-1}$, we have that
$N_f\leq 2^{n-1}-2^{n/2-1}-2^{m+1}$ for every $m$-resilient function $f$ with $m\leq n/2-2$.
 In the case  $n$ is odd, $N_f$ is smaller than or
equal to the highest multiple of $2^{m+1}$, which is less than or equal to the
best possible nonlinearity of all Boolean functions. In the sequel, we shall call ``Sarkar et al.'s bounds'' all these bounds.
We shall also extend the definitions of correlation-immune and resilient functions, so that our results are as general as possible: by convention, we shall say that any Boolean function is $0$th-order correlation immune and $(-1)$-resilient and that any balanced function is $0$-resilient.

We call   $(n,m)$-functions the functions  from  ${\Bbb F}_{2}^{n}$ to  ${\Bbb
F}_{2}^{m}$. Such function $F$ being given, the
Boolean functions $f_1,\ldots,f_m$ defined, at every $x\in  {\Bbb
F}_{2}^{n}$, by $F(x)=(f_1,\ldots,f_m)$, are called the coordinate
functions of $F$.  Obviously, these functions include the
(single-output) Boolean functions which correspond to the case
$m=1$. Furthermore, for $m=n$, the function  $F(x)=(f_1,\ldots,f_n)$
is called a Boolean permutation if $F(x)$  is a bijective mapping
from ${\Bbb F}_2^n$ to ${\Bbb F}_2^n$.

 The original Maiorana-McFarland's (M-M) class of bent functions
  \cite{McFarland1973}
is the set of all the (bent) Boolean functions on ${\Bbb
F}^{2n}_2=\{(x,y),x,y\in {\Bbb F}_{2}^{n}\}$ of the form:
  $$f(x,y)=x\cdot \phi(y) \oplus g(y)$$
 where $\phi(y)=(\phi_1(y),\phi_2(y),\ldots,\phi_n(y))$ is any
permutation on ${\Bbb F}_{2}^{n}$ and $g\in B_n$.
 \begin{lemma}\label{lem1}
   For $x\in {\Bbb F}_2^n, y\in {\Bbb F}_2^n$, let $\phi_{i}(y)$, $1\leq i \leq n$, be
    an $n$-variable Boolean
   function, and $g(y)$ be any $n$-variable Boolean
   function.  A $2n$-variable Boolean
   function   $f(x,y)=x\cdot \phi(y) \oplus g(y)=\bigoplus_{i=1}^n \phi_i(y_1,\dots ,y_n)x_i\oplus g(y_1,\dots ,y_n)$ is a bent function if
   and only if
   $$\phi(y)=(\phi_1(y), \phi_{2}(y),\ldots,\phi_{n}(y))$$
   is a Boolean permutation.
   \end{lemma}

This property comes directly from the fact that any restriction of $f$ obtained by fixing $y$ is affine. We shall say that the coordinates of $x$ are ``affine''. In the next section, we shall use such functions in a different - but equivalent - form: $n$ will be the global number of variables (instead of $2n$) and the ``affine'' variables will  be $x_{1},\dots ,x_{n/2}$, that is, the functions will have the form $f(x_1,\dots ,x_n)=\bigoplus_{i=1}^{n/2} \phi_i(x_{n/2+1},\dots ,x_{n})\, x_{i}\oplus g(x_{n/2+1},\dots ,x_{n})$.

\section{Secondary constructions of bent functions}\label{3}
  In this section, we present secondary constructions of bent functions. Before that, we first recall the concept of complementary plateaued functions.
  It will play an important role in the following constructions.

  \begin{definition}\label{def complementary pl}\cite{Zheng1999}
   Let $p$ be a positive odd number and $g_1, g_2\in B_{p}$. Then
   $g_1$ and $g_2$ are said to be complementary $(p-1)$th-order
   plateaued functions in $p$ variables if they are $p$-variable $(p-1)$th-order
   plateaued functions, and satisfy the property that
   $W_{g_1}(\omega)=0$ if and only if $W_{g_2}(\omega)\neq 0$.
  \end{definition}

\begin{lemma}\label{complementary plateaued}\cite{Zheng1999}
 Let $n$ be a positive even number and $x=(x_1,x_2,\ldots,x_n)\in {\Bbb
 F}_2^n$. Then $f(x)$ is bent if and only if the two functions,
 $f(x_1,\ldots,x_{j-1},0,x_{j+1},$ $\ldots,x_n)$ and
  $f(x_1,\ldots,x_{j-1},1,x_{j+1},\ldots,x_n)$ are complementary $(n-2)$th-order
  plateaued functions in
 $n-1$ variables, where $j=1,2,\ldots,n$.
\end{lemma}

In \cite{C.Carlet2004-bis}, Carlet  designed a
  secondary
  construction of bent functions, often called the {\em indirect
  sum}:
  \begin{corollary} \label{cor2}\cite{C.Carlet2004-bis,Carlet2010book}
Let $x\in \Bbb{F}_2^n, \, y\in
\Bbb{F}_2^m$. Let~$f_1$ and $f_2$ be two $n$-variable bent functions ($n$ even)
and let~$g_1$ and $g_2$ be two $m$-variable bent functions ($m$
even). Define $$h(x,y)= f_1(x)\oplus g_1(y)\oplus (f_1\oplus
f_2)(x)\, (g_1\oplus g_2)(y).$$ Then $h$ is bent and its dual is obtained from
$\widetilde f_1,\widetilde f_2,\widetilde g_1$ and $\widetilde g_2$
by the same formula as $h$ is obtained from $f_1,f_2,g_1$ and $g_2$.
\end{corollary}

 This above secondary  construction was altered into constructions of resilient functions, see \cite{C.Carlet2004-bis}, which includes as a particular case the well-know
\emph{direct sum} \cite{Siegenthaler84}, that we recall: for $x\in {\Bbb F}_2^n$ and $y\in {\Bbb F}_2^m$, let $f(x)$ be an $n$-variable $t$-resilient function ($t\geq 0$) and $g(y)$ be an $m$-variable
$k$-resilient function ($k\geq 0$), then the function
\[h(x,y)=f(x)\oplus g(y)\]
is a $(t+k+1)$-resilient function in $n+m$ variables. The nonlinearity
 of $h(x,y)$ is equal to $2^nN_{g}+2^mN_{f}-2N_{f}N_{g}$.

In the present paper,  we first modify the indirect sum into a new construction of bent functions:

\begin{construction}\label{construction bent 2}
Let $n$ and $m$ be two positive even numbers. For
$X=(x_1,\ldots,x_{n})\in {\Bbb F}_2^{n}$ and  $Y=(y_1,\ldots,y_{m})\in
{\Bbb F}_2^{m}$, $x=(x_1,\ldots,x_{\mu-1},x_{\mu+1},\ldots,x_n)\in \Bbb{F}_2^{n-1},
 y=(y_1,\ldots,$ $y_{\rho-1},y_{\rho+1},\ldots,y_m)\in \Bbb{F}_2^{m-1}$,  let $f(X)$ be an $n$-variable bent function and $g(Y)$ an $m$-variable bent function. We consider the restrictions of $f$ equal to
$f_0(x)=f(x_1,\ldots,x_{\mu-1},0,x_{\mu+1},$ \\$\ldots,x_n)$,
$f_1(x)=$ $f(x_1,\ldots,x_{\mu-1},1,x_{\mu+1},\ldots,x_n)$ and of $g$ equal to
$g_0(y)=g(y_1,\ldots,y_{\rho-1},0,y_{\rho+1},$ $\ldots,y_m)$,
$g_1(y)=g(y_1,\ldots,y_{\rho-1},1,y_{\rho+1},\ldots,y_m)$, where
$\mu\in\{1,2,\ldots,n\}$, $\rho\in\{1,2,\ldots,m\}$ and we define:
$$h(x,y)= f_0(x)\oplus g_0(y)\oplus (f_0\oplus f_1)(x)\, (g_0\oplus
g_1)(y).$$
\end{construction}
This construction indeed provides bent functions:
\begin{theorem}\label{theorem 1}
Let  $f(X)\in B_n, g(Y)\in B_m$ and $h(x,y)\in B_{n+m-2}$ be defined as in Construction \ref{construction bent
  2}. Then $h$
is a bent function in $n+m-2$ variables. Further, the dual of $h$ is
obtained from
$\overline{f_0}(x)=\widetilde{f}(x_1,\ldots,x_{\mu-1},0,x_{\mu+1},$
$\ldots,x_n)$,
$\overline{f_1}(x)=\widetilde{f}(x_1,\ldots,x_{\mu-1},1,$ $x_{\mu+1},\ldots,x_n)$,
$\overline{g_0}(y)=\widetilde{g}(y_1,\ldots,y_{\rho-1},0,y_{\rho+1},\ldots,$
$y_m)$ and
$\overline{g_1}(y)=\widetilde{g}(y_1,\ldots,y_{\rho-1},$ $1,y_{\rho+1},\ldots,y_m)$,
by the same formula as $h$ is obtained from $f_0,f_1,g_0$ and $g_1$.
\end{theorem}
\begin{proof}
 According to Definition \ref{bent def}, the bentness of $h(x,y)$
 will be proved if we can show that  $W_h(a,b)=\pm 2^{(n+m-2)/2}$ for
 every $ a=(a_1,\ldots,a_{\mu-1},a_{\mu+1},\ldots,a_n)\in \Bbb{F}_2^{n-1}$ and
 $b=(b_1,\ldots,b_{\rho-1},b_{\rho+1},\ldots,b_m)\in \Bbb{F}_2^{m-1}$.
  As shown in \cite{C.Carlet2004-bis} for all Boolean functions, we have:
  \begin{equation}\label{equ W h}
  \begin{array}{rl}
   W_h(a,b)&\hspace{-0.3cm}=\hspace{-0.1cm}\sum\limits_{x\in \Bbb{F}_2^{n-1}}\sum\limits_{y\in
   \Bbb{F}_2^{m-1}}(-1)^{h(x,y)\oplus a\cdot x\oplus b\cdot y}\\
     &\hspace{-0.3cm}=\hspace{-0.1cm}\sum\limits_{x\in \Bbb{F}_2^{n-1}\atop f_0\oplus f_1=0}\sum\limits_{y\in
   \Bbb{F}_2^{m-1}}(-1)^{f_0(x)\oplus a\cdot x}(-1)^{g_0(y)\oplus b\cdot y}\\
   &\hspace{-0.3cm}+\sum\limits_{x\in \Bbb{F}_2^{n-1} \atop f_0\oplus f_1=1}\sum\limits_{y\in
   \Bbb{F}_2^{m-1}}(-1)^{f_0(x)\oplus a\cdot x}(-1)^{g_1(y)\oplus b\cdot y}\\

   &\hspace{-0.3cm}= \hspace{-0.1cm}W_{g_0}(b)\sum\limits_{x\in \Bbb{F}_2^{n-1}\atop f_0\oplus f_1=0}(-1)^{f_0(x)\oplus a\cdot
   x}
   \hspace{-0.1cm}+W_{g_1}(b)\sum\limits_{x\in \Bbb{F}_2^{n-1}\atop f_0\oplus f_1=1}(-1)^{f_0(x)\oplus a\cdot
   x}\\
   &\hspace{-0.3cm}= \hspace{-0.1cm}W_{g_0}(b)\hspace{-0.2cm}\sum\limits_{x\in \Bbb{F}_2^{n-1}}(-1)^{f_0(x)\oplus a\cdot
   x}\left( \frac{1+(-1)^{(f_0\oplus f_1)(x)}}{2}\right)\\
   &\hspace{-0.3cm}+\hspace{-0.05cm}W_{g_1}(b)\hspace{-0.2cm}\sum\limits_{x\in \Bbb{F}_2^{n-1}}(-1)^{f_0(x)\oplus a\cdot
   x}\left( \frac{1-(-1)^{(f_0\oplus f_1)(x)}}{2}\right)\\

   &\hspace{-0.3cm}=\hspace{-0.1cm}\frac{1}{2}W_{g_0}(b)\left[W_{f_0}(a)+W_{f_1}(a) \right]

   \hspace{-0.1cm}+\frac{1}{2}W_{g_1}(b)\left[W_{f_0}(a)-W_{f_1}(a) \right].
   \end{array}
  \end{equation}

 From Lemma \ref{complementary plateaued}, $f_0$ and $f_1$ are
 complementary $(n-2)$th-order plateaued functions in $n-1$ variables, $g_0$ and $g_1$ are
 complementary $(m-2)$th-order plateaued functions in $m-1$ variables. According to
 Definition \ref{def complementary pl} and Definition \ref{def plateaued},
it follows that  $W_h(a,b)=\pm 2^{(n+m-2)/2}$ for
 every $ a\in \Bbb{F}_2^{n-1},  b\in \Bbb{F}_2^{m-1}$.

   Next, we show that the dual of $h$ is obtained from $\overline{f_0},\overline{f_1},\overline{g_0}$
   and $\overline{g_1}$. We have:
\begin{equation}\label{equ W h0}
  \begin{array}{rl}
   &W_f(a_1,\ldots,a_{\mu-1},0,a_{\mu+1},\ldots,a_n)\\
   &=2^{\frac n2}(-1)^{\overline{f_0}(a)}\\
   &=\hspace{-0.2cm}\sum\limits_{x\in \Bbb{F}_2^{n-1}\atop x_\mu=0}(-1)^{f_0(x)\oplus a\cdot x}
   +\hspace{-0.2cm}\sum\limits_{x\in \Bbb{F}_2^{n-1}\atop x_\mu=1}(-1)^{f_1(x)\oplus a\cdot x}\\
   &=W_{f_0}(a)+W_{f_1}(a).
  \end{array}
  \end{equation}
Further,
 \begin{equation}\label{equ W h1}
  \begin{array}{l}
   W_f(a_1,\ldots,a_{\mu-1},1,a_{\mu+1},\ldots,a_n)\\
   =2^{\frac n2}(-1)^{\overline{f_1}(a)}\\
   =\hspace{-0.1cm}\sum\limits_{x\in \Bbb{F}_2^{n-1}\atop x_\mu=0}(-1)^{f_0(x)\oplus a\cdot x}
   -\hspace{-0.1cm}\sum\limits_{x\in \Bbb{F}_2^{n-1}\atop x_\mu=1}(-1)^{f_1(x)\oplus a\cdot x}\\
   =W_{f_0}(a)-W_{f_1}(a).
  \end{array}
  \end{equation}
  Combining Relations (\ref{equ W h}), (\ref{equ W h0}) and (\ref{equ W
  h1}), we have
\begin{displaymath}
  \begin{array}{rl}
   W_h(a,b)&=2^{\frac{n+m}{2}-2}
   \left(
   (-1)^{\overline{g_0}(b)}+(-1)^{\overline{g_1}(b)}\right)(-1)^{\overline{f_0}(a)}\\
   &+2^{\frac{n+m}{2}-2}
   \left(
   (-1)^{\overline{g_0}(b)}-(-1)^{\overline{g_1}(b)}\right)(-1)^{\overline{f_1}(a)}\\
   &=2^{\frac{n+m}{2}-1}(-1)^{\widetilde{h}(a,b)}.

  \end{array}
  \end{displaymath}
 According to the above equality, it follows that
\begin{displaymath}
  \begin{array}{rl}
  (-1)^{\widetilde{h}(a,b)}&=\frac{1}{2} \left(
   (-1)^{\overline{g_0}(b)}+(-1)^{\overline{g_1}(b)}\right)(-1)^{\overline{f_0}(a)}\\
   &+\frac{1}{2}\left(
   (-1)^{\overline{g_0}(b)}-(-1)^{\overline{g_1}(b)}\right)(-1)^{\overline{f_1}(a)}.
  \end{array}
  \end{displaymath}
  Then we have
 \begin{displaymath}
  \begin{array}{c}
  {\widetilde{h}(a,b)}=\overline{g_0}(b)\oplus{\overline{f_0}(a)}\oplus\left(
   \overline{g_0}(b)\oplus{\overline{g_1}(b)}\right)\left(
   \overline{f_0}(a)\oplus{\overline{f_1}(a)}\right).
  \end{array}
  \end{displaymath}
That is,
\begin{displaymath}
  \begin{array}{c}
  {\widetilde{h}(x,y)}=\overline{g_0}(y)\oplus{\overline{f_0}(x)}\oplus \left(
   \overline{g_0}(y)\oplus{\overline{g_1}(y)}\right)\left(
   \overline{f_0}(x)\oplus{\overline{f_1}(x)}\right).
  \end{array}
  \end{displaymath}
\end{proof}

\begin{remark}
{\em Without loss of generality (up to linear equivalence) let us take $\mu=\rho=n$. Let us denote $e=(0,\dots ,0,1)$.
For any $x$ and $y$, we have  $(g_0\oplus g_1)(y)=D_eg(y,0)$ where ``,'' denotes concatenation et $
D_eg$ is the derivative of $g$, defined as $D_eg(y,0)=g(y,0)\oplus g(y,1)$. Then $h(x,y)=f(x,0)\oplus g(y,0)$ if $D_eg(y,0)=0$ and $h(x,y)=f(x,1)\oplus g(y,0)$ if $D_eg(y,0)=1$. Hence, $h(x,y)=f(x,0)\oplus g(y,0)\oplus D_ef(x,0)D_eg(y,0)=f(x,D_eg(y,0))\oplus g(y,0)=f(x,0)\oplus g(y,D_ef(x,0)$. The derivative plays a role in a construction from \cite{Carlet-Yucas2005} (which has been generalized in \cite{Carlet AAECC}), but the present construction is clearly different since it builds $(n+m-2)$-variable functions from $n$-variable and $m$-variable ones.}
\end{remark}

\begin{remark}\label{remark 2}{\em Taking $h(x,y)= f_1(x)\oplus g_0(y)\oplus (f_0\oplus f_1)(x)\, (g_0\oplus
g_1)(y)$ or $h(x,y)= f_0(x)\oplus g_1(y)\oplus (f_0\oplus f_1)(x)\, (g_0\oplus
g_1)(y)$ or $h(x,y)= f_1(x)\oplus g_1(y)\oplus (f_0\oplus f_1)(x)\, (g_0\oplus
g_1)(y)$ gives three other bent functions; of course these functions correspond to applying Construction \ref{construction bent
  2} to functions affinely equivalent to $f$ and $g$.}
\end{remark}

In what follows, we analyze the properties of $h(x,y)$. Before
that, we first introduce a notation. \emph{The algebraic degree of
variable $x_i$ }in $f$, denoted by $\deg(f,x_i)$, is the number of
variables in the longest term of $f$ that contains $x_i$.

\begin{proposition}\label{theorem degree}
 Let $n~(>2)$ and $m~(>2)$ be two even numbers.   Let  $f(X)\in B_n,  g(Y)\in B_m$ and $h(x,y)\in B_{n+m-2}$  be defined as in Construction \ref{construction bent
  2}. Then $2\leq \deg(h) \leq \frac{n+m-2}{2}-1$.
\end{proposition}
 \begin{proof}
  Clearly, $2\leq \deg(h)$ since $h$ is bent. If $\deg(f)=2$ and
  $\deg(g)$ $=2$, then $\deg(h)=2$.\\
  According to the bentness of   $f(X)$ (resp. $g(Y)$), we have $\deg(f)\leq n/2$ (resp. $\deg(g)\leq m/2$). Further,
  we have $\deg(f_0\oplus f_1)\leq n/2-1$ (resp. $\deg(g_0\oplus g_1)\leq
  m/2-1$)  because $\deg(f(x)\oplus f(x\oplus a)\leq n/2-1$, where $a\in {\Bbb F}_2^n$. Thus,  from Construction \ref{construction bent 2},
  we have $\deg(h) \leq \frac{n+m-2}{2}-1$, the equality holds if
  and only if $\deg(f, x_\mu)=n/2$ and $\deg(g,y_\rho)=m/2$.
 \end{proof}

\begin{remark}\label{remark degree}
{\em   If $m=2$, then $g(Y)=y_1y_2\oplus l(y_1,y_2)$, where $l(y_1,y_2)$ is an affine function. By Construction  \ref{construction bent 2}, we have
$\deg(f_0)\leq \deg(h)\leq \deg(f)\leq (n+m-2)/2$.  From Proposition \ref{theorem degree}, the $(n+m-2)$-variable functions constructed
  by Construction \ref{construction bent 2} have algebraic degree
  not exceeding $(n+m-2)/2-1$  if $n>2$ and $m>2$.  Thus, they can not belong to the $PS^-$
  class, since all $n$-variable functions in $PS^-$ have algebraic degree
  $n/2$ exactly \cite{J. Dillon}. In addition, the constructed function $h$ has
  algebraic degree $2$ if and only if both $f$ and $g$ have
  algebraic degree $2$.}
\end{remark}

 Let us apply Construction \ref{construction bent 2} to M-M functions  $f(x)=\bigoplus\limits_{i=1}^{n/2}\phi_i(x_{n/2+1},\dots ,$ $x_{n})x_{i}\oplus u(x_{n/2+1},\dots ,x_{n})$ and $
g(y)=\bigoplus\limits_{j=1}^{m/2}\psi_j(y_{m/2+1},\dots ,y_{m})y_{i}\oplus v(y_{m/2+1},
\dots ,$ $y_{m})$,
  where $u(x_{n/2+1},\dots ,x_{n})$ is any Boolean function in $n/2$ variables and $ v(y_{m/2+1},
\dots ,y_{m})$ is any Boolean function in $m/2$ variables.
  We
deduce the following primary construction:

\begin{corollary}\label{construction bent nMM}
  Let $n$ and $m$ be two positive even numbers and $\mu\in \{1,\ldots,n/2\}$,
    $\rho\in \{1,\ldots,m/2\}$. For
  $x=(x_1,\ldots,x_{\mu-1},x_{\mu+1},\ldots,x_n)\in \Bbb{F}_2^{n-1},
   y=(y_1,\ldots,$ $y_{\rho-1},y_{\rho+1},\ldots,y_m)\in \Bbb{F}_2^{m-1}$, let
  $\phi(x_{n/2+1},\dots,x_{n})=\left( \phi_1,\ldots,\phi_{n/2}\right)$ be a
  Boolean permutation in ${n}/{2}$
  variables and
   $\psi(y_{m/2+1},\dots,y_{m})=\left( \psi_1,\ldots,\psi_{m/2}\right)$ a
   Boolean permutation in ${m}/{2}$
  variables.
  Then the $(n+m-2)$-variable function
  \begin{equation}\label{equ bent nMM}
  \begin{array}{rl}
  \hspace{-0.5cm}h(x,y)&=\bigoplus\limits_{i=1\atop i\neq
   \mu}^{n/2}\phi_i(x_{n/2+1},\dots ,x_{n})\, x_{i}

   \oplus \bigoplus\limits_{j=1\atop j\neq \rho}^{m/2}\psi_j(y_{1+m/2},\dots ,y_{m})\, y_{j}\\

   &\oplus\,
   \phi_{\mu}(x_{n/2+1},\dots ,x_{n})\psi_{\rho}(y_{1+m/2},\dots ,y_{m})\\
   &\oplus\, u(x_{n/2+1},\dots ,x_{n})\oplus v(y_{1+m/2},
\dots ,y_{m})
 \end{array}
\end{equation}

   is bent, where  $ u(x_{n/2+1},\dots,x_{n})\in B_{n/2},v(y_{m/2+1},
\dots, $ $y_{m})\in B_{m/2}$.
\end{corollary}

\begin{remark}\label{remark c bent nMM}
 {\em  The bent functions given by Corollary \ref{construction bent nMM}, have a form similar to those of M-M functions; indeed,  $ \phi_{\mu}(x_{n/2+1},\dots ,x_{n})\psi_{\rho}(y_{m/2+1},\dots ,y_{m})$ does not depend on the ``affine'' variables. There are cases where $h(x,y)$ is an $(n+m-2)$-variable M-M bent function; for instance when $ \phi_{\mu}=x_l$ and $\left( \phi_1,\ldots,\phi_{\mu-1},
   \phi_{\mu+1},\phi_{n/2}\right)$ is a Boolean permutation in $n/2-1$ variables,
  or $ \psi_{\rho}=y_t$ and $\left( \psi_1,\ldots,\right.$ $\left.\psi_{\rho-1},
   \psi_{\rho+1},\psi_{m/2}\right)$ is a Boolean permutation in $m/2-1$
   variables.  But the functions of Corollary \ref{construction bent nMM} are in general not M-M functions; the mapping: $$\Theta: (x_{n/2+1},\dots ,x_{n},y_{m/2+1},\dots ,y_{m})\mapsto $$
   $$\left(\phi_1(x_{n/2+1},\dots ,x_{n}),\dots, \phi_{\mu-1}(x_{n/2+1},\dots ,x_{n}),\right.$$
   $$\phi_{\mu+1}(x_{n/2+1},\dots ,x_{n}),\dots,\phi_{n/2}(x_{n/2+1},\dots ,x_{n}),$$
   $$ \psi_1(y_{m/2+1},\dots ,y_{m}),\dots, \psi_{\rho-1}(y_{m/2+1},\dots ,y_{m}),$$
   $$\left.\psi_{\rho+1}(y_{m/2+1},\dots ,y_{m}),\dots ,
    \psi_{m/2}(y_{m/2+1},\dots ,y_{m})\right)$$
  is not a permutation; it is even not a vectorial function with an equal number of input and output bits. \\In  \cite[Proposition 1]{C. Carlet2004} is introduced  a generalization of the M-M construction: let $s\geq r$ and
let $\Theta$ be any mapping from $\Bbb{F}_2^s$
to $\Bbb{F}_2^r$  such that, for every $a\in \Bbb{F}_2^r$, the set $\Theta^{-1}(a)$
is an $(n-2r)$-dimensional affine subspace of~$\Bbb{F}_2^s$ and let $g$ be any Boolean function on~$\Bbb{F}_2^s$ whose restriction to $\Theta^{-1}(a)$ is bent for every $a\in \Bbb{F}_2^r$, if $n>2r$ (no condition on~$g$
being imposed if $n=2r$, which corresponds to the original M-M construction), then
$x\cdot \Theta(y)\oplus g(y)$ is bent.  We can see that Corollary \ref{construction bent nMM} is in some cases a particular case of this
  general construction of bent functions with $s=(m+n)/2, r=(m+n-4)/2$ (this happens for instance when $\Theta$ is an affine mapping).  But, in general, it is not, since the condition ``$\Theta^{-1}(a)$  is an $(n-2r)$-dimensional affine subspace of~$\Bbb{F}_2^s$'' is not satisfied.}
\end{remark}

   According to Remark \ref{remark 2} and Corollary \ref{construction bent nMM}, we know that $h(x,y)\oplus  \phi_{\mu}(x_{n/2+1},$ $\dots ,x_{n})$, $h(x,y)\oplus  \psi_{\rho}(y_{m/2+1},\dots ,y_{m})$ and $h(x,y)\oplus  \phi_{\mu}(x_{n/2+1},\dots ,x_{n})\oplus \psi_{\rho}(y_{m/2+1},\dots ,y_{m})$ are also bent functions,
  where $h(x,y)$ are defined as Corollary \ref{construction bent nMM}.  Further, similarly to Corollary \ref{construction bent nMM}, we  are able to select
  $\mu\in \{1,\ldots,n/2\}$,
    $\rho\in \{\frac{m}{2}+1,\ldots,m\}$ or  $\mu\in \{\frac{n}{2}+1,\ldots,n\}$,
    $\rho\in \{1,\ldots,m/2\}$ or   $\mu\in \{\frac{n}{2}+1,\ldots,n\}$,
 $\rho\in \{\frac{m}{2}+1,\ldots,m\}$. This gives three primary constructions similar to that of Corollary \ref{construction bent nMM}. We can also apply Construction \ref{construction bent 2} using as initial functions two elements of the $PS_{ap}$ class of bent functions (introduced in \cite{J. Dillon} and recalled for instance in \cite{Carlet2010book}). Recall that the functions of this class are defined over ${\Bbb F}_{2^{n/2}}\times {\Bbb F}_{2^{n/2}}\sim {\Bbb F}_2^n$ as $f(x,y)=g(x/y)$ where $x,y\in {\Bbb F}_{2^{n/2}}$ and $g$ is balanced on ${\Bbb F}_{2^{n/2}}$, with the convention $x/0=0$.
 To define $f_0$ we need to restrict $f$ to a linear hyperplane $\{(x,y)\in {\Bbb F}_{2^{n/2}}\times {\Bbb F}_{2^{n/2}}\, |\, Tr^{n/2}_1(ax\oplus by)=0\}$ of ${\Bbb F}_2^n$, where $Tr^{n/2}_1$ is the absolute trace over ${\Bbb F}_{2^{n/2}}$ and $(a,b)\neq (0,0)$. We have $(f_0\oplus f_1)(x,y)=D_{(\alpha ,\beta)}f(x,y)$ for some $(\alpha ,\beta)\in {\Bbb F}_{2^{n/2}}\times {\Bbb F}_{2^{n/2}}$ such that $tr(a\alpha +b\beta)=1$.

  \begin{corollary}\label{construction bent by PSab}
   Let $n$ and $m$ be two  positive even numbers. We identify  ${\Bbb F}_2^{n/2}$ (resp. $ {\Bbb F}_2^{m/2}$) with the Galois field ${\Bbb F}_{2^{n/2}}$ (resp. ${\Bbb F}_{2^{m/2}}$).  Let  $\theta$ (resp. $\vartheta$) be a balanced function on ${\Bbb F}_{2^{n/2}}$ (resp. $ {\Bbb F}_{2^{m/2}}$). Let $(x,y)\in {\Bbb F}_{2^{n/2}}\times {\Bbb F}_{2^{n/2}}$, $(z,\tau)\in {\Bbb F}_{2^{m/2}}\times {\Bbb F}_{2^{m/2}}$, let $f(x,y)=\theta(\frac{x}{y})$ for $y\neq 0$, otherwise $f(x,y)=0$, let $g(z,\tau)=\theta(\frac{z}{\tau})$ for $\tau\neq 0$, otherwise $g(z,\tau)=0$.
  Let $f_0(x,y)$ (resp. $g_0(z,\tau)$) be the restriction of $f$ (resp. $g$) on $\{(x,y)\in {\Bbb F}_{2^{n/2}}\times {\Bbb F}_{2^{n/2}}|Tr^{n/2}_1(ax\oplus by)=0\}$ (resp. $\{(z,\tau)\in {\Bbb F}_{2^{n/2}}\times {\Bbb F}_{2^{m/2}}|Tr^{m/2}_1(cz\oplus d\tau)=0 \}$), where $(a,b)\neq (0,0)\in {\Bbb F}_{2^{n/2}}\times {\Bbb F}_{2^{n/2}}$, $(c,d)\neq (0,0)\in {\Bbb F}_{2^{m/2}}\times {\Bbb F}_{2^{m/2}}$. We take  $f_1(x,y)=f_0(x\oplus \alpha,y\oplus \beta)$, where $Tr^{n/2}_1(a\alpha\oplus b\beta)=1, (\alpha,\beta)\in {\Bbb F}_{2^{n/2}}\times {\Bbb F}_{2^{n/2}}$ and
 $g_1(z,\tau)=g_0(z\oplus u,\tau\oplus v)$, where $Tr^{m/2}_1(cu\oplus dv)=1, (u,v)\in {\Bbb F}_{2^{m/2}}\times {\Bbb F}_{2^{m/2}}$.
Then
   \begin{displaymath}\label{equ PSab}
\begin{array}{r}
\hspace{-0.15cm}h(x,y,z,\tau)\!=f_0(x,y)\oplus g_0(z,\tau)
\oplus (f_0\oplus f_1)(x,y)\, (g_0\oplus
g_1)(z,\tau)
  \end{array}
\end{displaymath}
is a bent function on ${\Bbb F}_{2^{n+m-2}}$.
   \end{corollary}

 Of course we could also apply Construction \ref{construction bent 2} using as initial functions an M-M function and a function of $PS_{ap}$.

In 1976,  Rothaus presented a secondary construction which
uses three initial $n$-variable bent functions $f^{(1)},f^{(2)},
f^{(3)}$ to build
a fourth one $f$ which is an $(n+2)$-variable bent function:\\
\textbf{Rothaus' construction} \cite{O.S.Rothaus}: Let
$x=(x_1,x_2,\ldots, x_n)\in {\Bbb F}_2^n$ and $x_{n+1},x_{n+2}\in
{\Bbb F}_2$. Let $f^{(1)}(x)$, $f^{(2)}(x)$, $f^{(3)}(x)$   be bent
functions on ${\Bbb F}_2^{n}$ such that  $f^{(1)}(x)\oplus
f^{(2)}(x) \oplus f^{(3)}(x)$ is bent as well, then the function
defined at every element $(x,x_{n+1},x_{n+2})\in {\Bbb F}_2^{n+2}$
by:
\begin{displaymath}\label{Rothaus}
\begin{array}{rl}
  f(x,x_{n+1},x_{n+2})\!\!&=f^{(1)}(x)f^{(2)}(x)\oplus f^{(1)}(x)f^{(3)}(x)\\
 & \oplus f^{(2)}(x)f^{(3)}(x)
  \oplus [f^{(1)}(x)\oplus f^{(2)}(x)]x_{n+1}\\
 & \oplus [f^{(1)}(x)\oplus
  f^{(3)}(x)]x_{n+2}\oplus x_{n+1}x_{n+2}
  \end{array}
\end{displaymath}
is a bent function in $n+2$ variables.

We apply Construction \ref{construction bent 2} to bent functions constructed by Rothaus' construction.
   \begin{corollary}\label{construction bent Rothaus}
   Let $n$ and $m$ be two positive even numbers and $x\in {\Bbb F}_2^n, y\in {\Bbb F}_2^m,$ $x_{n+1},x_{n+2},y_{m+1},y_{m+2}\in
{\Bbb F}_2$. Let an $(n+2)$-variable bent function $f$
    and an $(m+2)$-variable bent function $g$ be built by means of Rothaus' construction, respectively from $n$-variable  bent functions $f^{(1)},f^{(2)},
f^{(3)}$ and $m$-variable  bent functions $g^{(1)},g^{(2)},
g^{(3)}$.
Then
   \begin{equation}\label{equ Rothaus}
\begin{array}{rl}
\hspace{-0.15cm}h(x,y,x_{n+1},y_{m+1}\!)\!&=\!\!f^{(1)}(x)f^{(2)}(x)\oplus f^{(1)}(x)f^{(3)}(x)\\
&\oplus
  f^{(2)}(x)f^{(3)}(x)\oplus g^{(1)}(y)g^{(2)}(y)\\
 &\oplus g^{(1)}(y)g^{(3)}(y)\oplus
  g^{(2)}(y)g^{(3)}(y)\\
  &\oplus [f^{(1)}(x)\oplus f^{(2)}(x)]x_{n+1}\\
 &\oplus [g^{(1)}(y)\oplus g^{(2)}(y)]y_{m+1}\\
  &\oplus [f^{(1)}\!(x)\!\oplus\!
  f^{(3)}\!(x)][g^{(1)}\!(y)\!\oplus\!
  g^{(3)}\!(y)]\\
   &\oplus [f^{(1)}(x)\oplus f^{(3)}(x)]y_{m+1}\\
 &\oplus [g^{(1)}(y)\oplus g^{(3)}(y)]x_{n+1}\\
  &\oplus x_{n+1} y_{m+1}.
  \end{array}
\end{equation}
is a bent function in $n+m+2$ variables.
   \end{corollary}
   \begin{proof}
   We select $f$ and $g$ as the initial functions of Construction \ref{construction bent 2} and set $\mu=n+2, \rho=m+2$. From  Theorem \ref{theorem 1}, we know that $h(x,y,x_{n+1},y_{m+1})$  is a bent function in $n+m+2$ variables.
   \end{proof}

 Next, we consider the bent functions in class $D$ as the initial functions of Construction \ref{construction bent 2}. We first introduce class $D$, which  has been derived in \cite{C. Carlet1994} from M-M bent functions, by adding to some functions of this class the indicators of some vector subspaces:\\
 The class $D$ of all the functions of the form $\bigoplus\limits_{i=1}^{n/2}\phi_i(x_{n/2+1},\dots ,x_{n})x_{i}\oplus 1_{E_1}(x_1,$ $\dots, x_{n/2})1_{E_2}(x_{n/2+1},$ $ \dots,x_{n}),$  where $\phi$ is any permutation on  ${\Bbb F}_2^{n/2}$,
         $E_1, E_2$ are two linear subspaces of  ${\Bbb F}_2^{n/2}$ such that $\phi(E_2)=E_1^{\perp}$ and $1_{E_1}(x_1,\dots, x_{n/2})$ (resp. $1_{E_2}(x_{n/2+1}, \dots,x_{n}$) is the characteristic function of $E_1$ (resp. $E_2$).

  \begin{corollary}\label{construction bent D}
  Let $n$ and $m$ be two positive even numbers and $\mu\in \{1,\ldots,n/2\}$,
    $\rho\in \{1,\ldots,m/2\}$. For
  $x=(x_1,\ldots,x_{\mu-1},x_{\mu+1},\ldots,x_n)\in \Bbb{F}_2^{n-1},
   y=(y_1,\ldots,$ $y_{\rho-1},y_{\rho+1},\ldots,y_m)\in \Bbb{F}_2^{m-1}$, let
  $\phi(X_1^{(n/2)})=\left( \phi_1,\ldots,\phi_{n/2}\right)$ be a
  Boolean permutation in $\frac{n}{2}$
  variables and
   $\psi(Y_1^{(m/2)})=\left( \psi_1,\ldots,\psi_{m/2}\right)$ a
   Boolean permutation in $\frac{m}{2}$
  variables,  where $X_1^{(n/2)}=(x_{n/2+1},\dots,x_{n}),$ $ Y_1^{(m/2)}=(y_{m/2+1},\dots,y_{m})$.   Let
  $E_1, E_2$ (resp. $\Xi_1, \Xi_2$) be two linear subspaces of  ${\Bbb F}_2^{n/2}$ (resp. ${\Bbb F}_2^{m/2}$) such that $\phi(E_2)=E_1^{\perp}$ (resp. $\psi(\Xi_2)=\Xi_1^{\perp}$).
  Then the $(n+m-2)$-variable function
  \begin{displaymath}
  \begin{array}{rl}
  \hspace{-0.2cm}h(x,y)=&\hspace{-0.2cm}\bigoplus\limits_{i=1\atop i\neq
   \mu}^{n/2}\phi_i( X_1^{(n/2)})\, x_{i}\oplus \bigoplus\limits_{j=1\atop j\neq \rho}^{m/2}\psi_j( Y_1^{(m/2)})\, y_{j}\\
   \end{array}
\end{displaymath}
 \begin{displaymath}
  \begin{array}{rl}
   &\hspace{-0.6cm}\oplus \bigoplus\limits_{\tau\in E_1}(\tau_{\mu}\oplus 1)\prod\limits_{i=1 \atop i \neq \mu}^{n/2}(x_i\oplus \tau_i\oplus 1)1_{E_2}( X_1^{(n/2)})\\

   &\hspace{-0.6cm} \oplus \bigoplus\limits_{\varsigma\in \Xi_1}(\varsigma_{\rho}\oplus 1)\prod\limits_{j=1 \atop j \neq \rho}^{m/2}(y_j\oplus \varsigma_j\oplus 1) 1_{\Xi_2}( Y_1^{(m/2)})\\

   &\hspace{-0.6cm}
   \oplus\,
   \phi_{\mu}( X_1^{(n/2)})\psi_{\rho}( Y_1^{(m/2)})\\
   &\hspace{-0.6cm}\oplus \psi_{\rho}( Y_1^{(m/2)})\left( \bigoplus\limits_{\tau\in E_1}\prod\limits_{i=1 \atop i \neq \mu}^{n/2}(x_i\oplus \tau_i\oplus 1)\right)1_{E_2}( X_1^{(n/2)})\\
   &\hspace{-0.6cm}\oplus \phi_{\mu}( X_1^{(n/2)})\left( \bigoplus\limits_{\varsigma\in \Xi_1}\prod\limits_{j=1 \atop j \neq \rho}^{m/2}(y_j\oplus \varsigma_j\oplus 1)\right)1_{\Xi_2}( Y_1^{(m/2)})\\
   &\hspace{-0.6cm}\oplus \left( \bigoplus\limits_{\tau\in E_1}\prod\limits_{i=1 \atop i \neq \mu}^{n/2}(x_i\oplus \tau_i\oplus 1)\right)1_{E_2}( X_1^{(n/2)})\\
   &\hspace{-0.6cm}\times\left( \bigoplus\limits_{\varsigma\in \Xi_1}\prod\limits_{j=1 \atop j \neq \rho}^{m/2}(y_j\oplus \varsigma_j\oplus 1)\right)1_{\Xi_2}( Y_1^{(m/2)}).

   \end{array}
  \end{displaymath}
   is bent.
\end{corollary}
\begin{proof}
   Let $f=\bigoplus\limits_{i=1}^{n/2}\phi_i( X_1^{(n/2)})\, x_{i}\oplus 1_{E_1}(x_1,\dots,$ $ x_{n/2})1_{E_2}(X_1^{(n/2)}), g=\bigoplus\limits_{j=1}^{m/2}\psi_j( Y_1^{(m/2)})\, y_{j} \oplus 1_{\Xi_1}(y_1,\dots,$ $ y_{m/2})1_{\Xi_2}(Y_1^{(m/2)})$. Clearly, $h(x,y)$
    is a bent function in $n+m-2$ variables if we select $f$ and $g$ as the initial functions of Construction \ref{construction bent 2}.
   \end{proof}

 \section{Secondary constructions of highly nonlinear functions}\label{4}

  In this section, we present a generalization of the indirect sum construction for constructing resilient functions with high nonlinearity. Before that, we first recall the secondary construction of bent functions deduced by  Carlet, Zhang and Hu in \cite{Carlet2012Zhang}.
  \begin{lemma}\label{CZHconstruction2}
   Let $n$ and $m$ be two even positive integers. Let $f_1(x),
   f_2(x)$ and $f_3(x)$ be bent functions in $n$ variables.
   Let $g_1(y),
   g_2(y)$ and $g_3(y)$ be bent functions in $m$ variables.
   Denote by $\nu_1$ the function $f_1\oplus f_2\oplus f_3$ and by
   $\nu_2$ the function $g_1\oplus g_2 \oplus g_3$. If both $\nu_1$ and
   $\nu_2$ are bent functions and if
    $\widetilde{\nu_1}=\widetilde{f_1}\oplus \widetilde{f_2}\oplus
    \widetilde{f_3}$,
    then
    \[f(x,y)=f_1(x)\oplus g_1(y)\oplus (f_1\oplus f_2)(x)(g_1\oplus g_2)(y)
    \oplus (f_2\oplus f_3)(x)(g_2\oplus g_3)(y)\]
    is a bent function in $n+m$ variables.
  \end{lemma}

   Now, we  adapt the above construction for constructing resilient functions.

   \begin{theorem}\label{the 4.1}
   Let $n$,    $m$, $t$  and $k$ be four integers such that $-1\leq t<n$ and $-1\leq k<m$. Let $f_1(x),
   f_2(x)$ and $f_3(x)$ be three $t$-resilient  functions in $n$ variables.
   Let $g_1(y),
   g_2(y)$ and $g_3(y)$ be $k$-resilient functions in $m$ variables. If $f_1(x)\oplus
   f_2(x)\oplus f_3(x)$ is also a $t$-resilient  function in $n$ variables and   $g_1(y)\oplus
   g_2(y)\oplus g_3(y)$ is also an $r$-resilient  function in $m$ variables,
    then the function
   \[\begin{array}{rl}f(x,y)\hspace{-0.1cm}=f_1(x)\oplus g_1(y)\oplus (f_1\oplus f_2)(x)(g_1\oplus g_2)(y)
    \oplus (f_2\oplus f_3)(x)(g_2\oplus g_3)(y)\end{array}\]
    is a $(t+k+1)$-resilient function in $n+m$ variables.

   \end{theorem}

   \begin{proof}
      From Lemma \ref{xiao messay}, $f(x,y)$ is a $(t+k+1)$-resilient function in $n+m$ variables if we can prove that $W_f(\alpha,\beta)$ is null for every
      $\alpha\in {\Bbb F}_2^n, \beta\in {\Bbb F}_2^m$ such that $0\leq wt(\alpha,\beta)\leq t+k+1$. We have:
       \begin{equation}\label{equ resilient}
  \begin{array}{rl}
   &W_f(\alpha,\beta)\\
   &\hspace{-0.2cm}=\sum\limits_{x\in \Bbb{F}_2^{n}}\sum\limits_{y\in
   \Bbb{F}_2^{m}}(-1)^{f(x,y)\oplus \alpha\cdot x\oplus \beta\cdot y}\\

     &\hspace{-0.2cm}=\sum\limits_{x\in \Bbb{F}_2^{n},\atop f_1(x)=f_2(x)=f_3(x)=0}(-1)^{ \alpha\cdot x}\sum\limits_{y\in
   \Bbb{F}_2^{m}}(-1)^{g_1(y)\oplus \beta\cdot y}\\

  &\hspace{-0.2cm}+ \hspace{-0.2cm}\sum\limits_{x\in \Bbb{F}_2^{n},\atop f_1(x)=f_2(x)=f_3(x)=1}(-1)^{ 1\oplus\alpha\cdot x}\sum\limits_{y\in
   \Bbb{F}_2^{m}}(-1)^{g_1(y)\oplus \beta\cdot y}\\

    &\hspace{-0.2cm}+\hspace{-0.2cm}\sum\limits_{x\in \Bbb{F}_2^{n},\atop f_1(x)\neq f_2(x)=f_3(x)=0}\hspace{-0.4cm}(-1)^{ 1\oplus \alpha\cdot x}\sum\limits_{y\in
   \Bbb{F}_2^{m}}(-1)^{g_2(y)\oplus \beta\cdot y}\\

   &\hspace{-0.2cm}+\hspace{-0.2cm}\sum\limits_{x\in \Bbb{F}_2^{n},\atop f_1(x)\neq f_2(x)=f_3(x)=1}\hspace{-0.4cm}(-1)^{\alpha\cdot x}\sum\limits_{y\in
   \Bbb{F}_2^{m}}(-1)^{g_2(y)\oplus \beta\cdot y}\\

    &\hspace{-0.2cm}+\hspace{-0.2cm}\sum\limits_{x\in \Bbb{F}_2^{n},\atop f_2(x)\neq f_1(x)=f_3(x)=0}\hspace{-0.4cm}(-1)^{ \alpha\cdot x}\sum\limits_{y\in
   \Bbb{F}_2^{m}}(-1)^{g_3(y)\oplus \beta\cdot y}\\
\end{array}
  \end{equation}

    \begin{displaymath}
  \begin{array}{rl}

    &\hspace{-0.2cm}+\hspace{-0.2cm}\sum\limits_{x\in \Bbb{F}_2^{n},\atop f_2(x)\neq f_1(x)=f_3(x)=1}\hspace{-0.4cm}(-1)^{ 1\oplus \alpha\cdot x}\sum\limits_{y\in
   \Bbb{F}_2^{m}}(-1)^{g_3(y)\oplus \beta\cdot y}\\

    &\hspace{-0.2cm}+\hspace{-0.2cm}\sum\limits_{x\in \Bbb{F}_2^{n},\atop f_3(x)\neq f_1(x)=f_2(x)=0}\hspace{-0.4cm}(-1)^{ \alpha\cdot x}\sum\limits_{y\in
   \Bbb{F}_2^{m}}(-1)^{g_1(y)\oplus g_2(y)\oplus g_3(y)\oplus \beta\cdot y}\\

   &\hspace{-0.2cm}+\hspace{-0.2cm}\sum\limits_{x\in \Bbb{F}_2^{n},\atop f_3(x)\neq f_1(x)=f_2(x)=1}\hspace{-0.5cm}(-1)^{1\oplus  \alpha\cdot x}\hspace{-0.2cm}\sum\limits_{y\in
   \Bbb{F}_2^{m}}(-1)^{g_1(y)\oplus g_2(y)\oplus g_3(y)\oplus \beta\cdot y}\\

   &\hspace{-0.2cm}= W_{g_1}(\beta)\left[\sum\limits_{x\in \Bbb{F}_2^{n},\atop f_1(x)=f_2(x)=f_3(x)=0}\hspace{-0.5cm}(-1)^{ \alpha\cdot
   x}-\hspace{-0.4cm}\sum\limits_{x\in \Bbb{F}_2^{n},\atop f_1(x)=f_2(x)=f_3(x)=1}\hspace{-0.7cm}(-1)^{\alpha\cdot
   x}\right]\\

   &\hspace{-0.2cm}+ W_{g_2}(\beta)\left[\sum\limits_{x\in \Bbb{F}_2^{n},\atop f_1(x)\neq f_2(x)=f_3(x)=1}\hspace{-0.5cm}(-1)^{ \alpha\cdot
   x}-\hspace{-0.4cm}\sum\limits_{x\in \Bbb{F}_2^{n}, \atop f_1(x)\neq f_2(x)=f_3(x)=0}\hspace{-0.5cm}(-1)^{ \alpha\cdot
   x}\right]\\

   &\hspace{-0.2cm}+ W_{g_3}(\beta)\left[\sum\limits_{x\in \Bbb{F}_2^{n}\atop f_2(x)\neq f_1(x)=f_3(x)=0}\hspace{-0.5cm}(-1)^{ \alpha\cdot
   x}-\hspace{-0.4cm}\sum\limits_{x\in \Bbb{F}_2^{n}\atop f_2(x)\neq f_1(x)=f_3(x)=1}\hspace{-0.5cm}(-1)^{ \alpha\cdot
   x}\right]\\

   &\hspace{-0.2cm}+ W_{g_1\oplus g_2\oplus g_3}(\beta)\left[\sum\limits_{x\in \Bbb{F}_2^{n} \atop  {f_1(x)=f_2(x)=0 \atop f_3(x)=1}}\hspace{-0.5cm}(-1)^{ \alpha\cdot
   x}-\hspace{-0.4cm}\sum\limits_{x\in \Bbb{F}_2^{n} \atop {f_1(x)=f_2(x)=1 \atop f_3(x)=0}}\hspace{-0.5cm}(-1)^{ \alpha\cdot
   x}\right]\\

    &\hspace{-0.2cm}= W_{g_1}(\beta)\left[\sum\limits_{x\in \Bbb{F}_2^{n}}(-1)^{ \alpha\cdot
   x}( \frac{1+(-1)^{f_1(x)}}{2})( \frac{1+(-1)^{f_2(x)}}{2})\right.\\
   &( \frac{1+(-1)^{f_3(x)}}{2})
   -\sum\limits_{x\in \Bbb{F}_2^{n}}(-1)^{\alpha\cdot
   x}( \frac{1-(-1)^{f_1(x)}}{2})( \frac{1-(-1)^{f_2(x)}}{2})\\
   & \left.( \frac{1-(-1)^{f_3(x)}}{2})\right]\\

   &\hspace{-0.2cm}+ W_{g_2}(\beta)\left[\sum\limits_{x\in \Bbb{F}_2^{n}}(-1)^{ \alpha\cdot
   x}( \frac{1+(-1)^{f_1(x)}}{2})( \frac{1-(-1)^{f_2(x)}}{2})\right.\\
   &( \frac{1-(-1)^{f_3(x)}}{2})
    -\sum\limits_{x\in \Bbb{F}_2^{n}}(-1)^{\alpha\cdot
   x}( \frac{1-(-1)^{f_1(x)}}{2})( \frac{1+(-1)^{f_2(x)}}{2})\\
   &\left.( \frac{1+(-1)^{f_3(x)}}{2})\right]\\

   &\hspace{-0.2cm}+ W_{g_3}(\beta)\left[\sum\limits_{x\in \Bbb{F}_2^{n}}(-1)^{ \alpha\cdot
   x}( \frac{1+(-1)^{f_1(x)}}{2})( \frac{1-(-1)^{f_2(x)}}{2})\right.\\
   &( \frac{1+(-1)^{f_3(x)}}{2})
   -\sum\limits_{x\in \Bbb{F}_2^{n}}(-1)^{\alpha\cdot
   x}( \frac{1-(-1)^{f_1(x)}}{2})( \frac{1+(-1)^{f_2(x)}}{2})\\
    &\left.( \frac{1-(-1)^{f_3(x)}}{2})\right]\\

   &\hspace{-0.2cm}+ W_{g_1\oplus g_2\oplus g_3}(\beta)\left[\sum\limits_{x\in \Bbb{F}_2^{n}}(-1)^{ \alpha\cdot
   x}( \frac{1+(-1)^{f_1(x)}}{2})( \frac{1+(-1)^{f_2(x)}}{2})\right.\\

   &( \frac{1-(-1)^{f_3(x)}}{2})
    -\sum\limits_{x\in \Bbb{F}_2^{n}}(-1)^{\alpha\cdot
   x}( \frac{1-(-1)^{f_1(x)}}{2})( \frac{1-(-1)^{f_2(x)}}{2})\\
   &\left.( \frac{1+(-1)^{f_3(x)}}{2})\right]
   \end{array}
  \end{displaymath}

Hence:

\begin{equation}\label{Walsh-th2}
  \begin{array}{rl}  W_f(\alpha,\beta)&\hspace{-0.2cm}=\frac{1}{4}W_{g_1}(\beta)\left[ W_{f_1}(\alpha)+W_{f_2}(\alpha)\right.
  \left.+W_{f_3}(\alpha)+W_{f_1\oplus f_2\oplus f_3}(\alpha)\right]\\

   &\hspace{-0.2cm}  +\frac{1}{4}W_{g_2}(\beta)\left[ W_{f_1}(\alpha)-W_{f_2}(\alpha)\right.\left.-W_{f_3}(\alpha)+W_{f_1\oplus f_2\oplus f_3}(\alpha)\right]\\

   &\hspace{-0.2cm}  +\frac{1}{4}W_{g_3}(\beta)\left[ W_{f_1}(\alpha)-W_{f_2}(\alpha)\right.\left.+W_{f_3}(\alpha)-W_{f_1\oplus f_2\oplus f_3}(\alpha)\right]\\

   &\hspace{-0.2cm} +\frac{1}{4}W_{g_1\oplus g_2\oplus g_3}(\beta)\left[ W_{f_1}(\alpha)+W_{f_2}(\alpha)\right.\left.-W_{f_3}(\alpha)-W_{f_1\oplus f_2\oplus f_3}(\alpha)\right].

  \end{array}
  \end{equation}
Since  $f_1,
   f_2$, $f_3$ and $f_1\oplus
   f_2\oplus f_3$ are $t$-resilient,  we have
   $W_{f_i}(\alpha)=0$ and $W_{f_1\oplus
   f_2\oplus f_3}(\alpha)=0$  for  any
      $ \alpha\in {\Bbb F}_2^n$ such that $0\leq wt(\alpha)\leq t$ , where $i=1,2,3$.  Since $g_1,
   g_2$, $g_3$ and $g_1\oplus
   g_2\oplus g_3$ are $k$-resilient,  we have $W_{g_i}(\beta)=0$ and $W_{g_1\oplus
   g_2\oplus g_3}(\beta)=0$  for  any
      $ \beta\in {\Bbb F}_2^m$ such that $0\leq wt(\beta)\leq k$, where $i=1,2,3$.
      In addition, we have $wt(\alpha)\leq t$ or $wt(\beta)\leq k$
      if $wt(\alpha,\beta)\leq t+k+1$. Further, according to
      Relation (\ref{equ resilient}), $f(x,y)$ is a  $(t+k+1)$-resilient function in $n+m$ variables.
   \end{proof}

   \begin{remark}
     {\em The indirect sum is a particular case of this construction: it  corresponds
      to the case $ f_2=f_3$ and $g_2=g_3$.}
      \end{remark}

We modify now the construction of Theorem \ref{the 4.1} to ensure a high nonlinearity of the constructed resilient function: to this aim, we assume that the functions $f_i$ are bent (of course, they can then not be balanced and the order $t$ of Theorem \ref{the 4.1} is then equal to $-1$). Before that, we first present a lemma.
 \begin{lemma}\label{lem increase}
   Let $n~ (>6)$  be an even positive integer and $m$  be  a positive integer. Let $f_1(x),
   f_2(x)$ and $f_3(x)$ be bent functions in $n$ variables such that $\nu_1=f_1\oplus f_2\oplus f_3$ is a bent function and
    $\widetilde{\nu_1}=\widetilde{f_1}\oplus \widetilde{f_2}\oplus
    \widetilde{f_3}$.
   Let $g_1(y),
   g_2(y)$ and $g_3(y)$ be  functions in $m$ variables.
   Denote by
   $\nu_2$ the function $g_1\oplus g_2 \oplus g_3$.  Let $f(x,y)$ be defined as in Theorem \ref{the 4.1} and  $\alpha\in {\Bbb F}_2^n, \beta\in {\Bbb F}_2^m$. Then, there are four cases.
  \begin{enumerate}
    \item If $W_{f_1}(\alpha)=W_{f_2}(\alpha)=W_{f_3}(\alpha)$, then
         $W_{\nu_1}(\alpha)=W_{f_1}(\alpha)$. Further,
         \[ W_f(\alpha,\beta)=W_{g_1}(\beta) W_{f_1}(\alpha); \]
    \item  If $W_{f_1}(\alpha)=W_{f_2}(\alpha)\neq W_{f_3}(\alpha)$, then
         $W_{\nu_1}(\alpha)=W_{f_3}(\alpha)$. Further,
         \[ W_f(\alpha,\beta)=W_{g_1\oplus g_2\oplus g_3}(\beta) W_{f_1}(\alpha); \]
    \item If $W_{f_1}(\alpha)\neq W_{f_2}(\alpha)=W_{f_3}(\alpha)$, then
         $W_{\nu_1}(\alpha)=W_{f_1}(\alpha)$. Further,
         \[ W_f(\alpha,\beta)=W_{g_2}(\beta) W_{f_1}(\alpha); \]
    \item If $W_{f_1}(\alpha)=W_{f_3}(\alpha)\neq W_{f_2}(\alpha)$, then
         $W_{\nu_1}(\alpha)=W_{f_2}(\alpha)$. Further,
         \[ W_f(\alpha,\beta)=W_{g_3}(\beta) W_{f_1}(\alpha).\]
  \end{enumerate}
 \end{lemma}
 \begin{proof}
 Since $\nu_1(x)$ is a bent function in $n$ variables and  $\widetilde{\nu_1}=\widetilde{f_1}\oplus \widetilde{f_2}\oplus
    \widetilde{f_3}$, then
    \[(-1)^{\widetilde{f_1}\oplus \widetilde{f_2}\oplus
    \widetilde{f_3}}=(-1)^{\widetilde{\nu_1}},\]
    that is,
    \begin{equation}\label{equ th1}
    \begin{array}{c}
     W_{f_1}(\alpha)W_{f_2}(\alpha)W_{f_3}(\alpha)=2^n W_{\nu_1}(\alpha).
     \end{array}
     \end{equation}

     We also know that $ W_{f_i}(\alpha)=\pm 2^{n/2}$ for any $\alpha\in {\Bbb F}_2^n$, where $i=1,2,3$. Thus, combining Relations (\ref{Walsh-th2}) and  (\ref{equ th1}),  the conclusion is held.

 \end{proof}

     \begin{theorem}\label{the bent A 4.1}
   Let $n~ (>6)$  be an even positive integer. Let  $m$  and $k$ be two integers such that
    $ k<m-1$. Let $f_1(x),
   f_2(x)$ and $f_3(x)$ be bent functions in $n$ variables.
   Let $g_1(y),
   g_2(y)$ and $g_3(y)$ be $k$-resilient functions in $m$ variables.
   Denote by $\nu_1$ the function $f_1\oplus f_2\oplus f_3$ and by
   $\nu_2$ the function $g_1\oplus g_2 \oplus g_3$. If  $\nu_1$ is a bent function,
   $\nu_2$ is a $k$-resilient function  and if
    $\widetilde{\nu_1}=\widetilde{f_1}\oplus \widetilde{f_2}\oplus
    \widetilde{f_3}$,
    then
    \[\begin{array}{c}f(x,y)=f_1(x)\oplus g_1(y)\oplus (f_1\oplus f_2)(x)(g_1\oplus g_2)(y)\\
    \oplus (f_2\oplus f_3)(x)(g_2\oplus g_3)(y)\end{array}\]
    is a $k$-resilient function in $n+m$ variables. Further, we have
   { \begin{equation}\label{nonlinearity-th2}
      \begin{array}{c}
         N_{f}\geq 2^{n+m-1}-
         2^{n/2-1}\times\max\left\{\max\limits_{\beta\in {\Bbb F}_2^m}\{|W_{g_1}(\beta)|\}, \right. \\
          \left. \max\limits_{\beta\in {\Bbb F}_2^m}\{|W_{g_2}(\beta)|\},\max\limits_{\beta\in {\Bbb F}_2^m}\{|W_{g_3}(\beta)|\}, \max\limits_{\beta\in {\Bbb F}_2^m}\{|W_{\nu_2}(\beta)|\}\right\};
      \end{array}
    \end{equation}
    and the equality holds if and only if $\{f_1, f_1\oplus 1\} \cap \{f_2, f_2\oplus 1\}=
     (\{f_1, f_1\oplus 1\} \cap \{f_3, f_3\oplus 1\})=(\{f_2, f_2\oplus 1\}
     \cap \{f_3, f_3\oplus 1\})=\emptyset$.}
   \end{theorem}

   \begin{proof}
According to Theorem \ref{the 4.1}, $f(x,y)$ is a  $k$-resilient function in $n+m$ variables.

    Next, we consider the nonlinearity of $f(x,y)$.
      From Lemma \ref{lem increase}, we immediately have
      \begin{displaymath}
      \begin{array}{c}
         N_{f}\geq 2^{n+m-1}-
         2^{n/2-1}\times\max\left\{\max\limits_{\beta\in {\Bbb F}_2^m}\{|W_{g_1}(\beta)|\}, \right. \\
          \left. \max\limits_{\beta\in {\Bbb F}_2^m}\{|W_{g_2}(\beta)|\},\max\limits_{\beta\in {\Bbb F}_2^m}\{|W_{g_3}(\beta)|\}, \max\limits_{\beta\in {\Bbb F}_2^m}\{|W_{\nu_2}(\beta)|\}\right\},
      \end{array}
    \end{displaymath}
    the equality holds if and only if  all four cases of Lemma \ref{lem increase} can happen, that is,
    $\{f_1, f_1\oplus 1\} \cap \{f_2, f_2\oplus 1\}=
     (\{f_1, f_1\oplus 1\} \cap \{f_3, f_3\oplus 1\})=(\{f_2, f_2\oplus 1\} \cap \{f_3, f_3\oplus 1)=\emptyset$.
   \end{proof}
   \begin{remark}
   {\em Theorem \ref{the bent A 4.1} allows constructing resilient functions
    offering a compromize between resiliency order (whose ratio
    with the number of variables is lowered when we move from
    functions $g_i$ to $f$) and nonlinearity (which is enhanced thanks to the
    contribution of the bent functions, resulting in the coefficient $2^{n/2-1}$ in
    Relation (\ref{nonlinearity-th2})). This is useful cryptographically speaking since low order resilient functions with high nonlinearity are more useful than high order resilient functions (with inevitably low nonlinearity according to the Sarkar-Maitra bound). If the nonlinearity of $m$-variable resilient functions $g_1,g_2,g_3$ and $g_1\oplus g_2\oplus g_3$ can exceed $2^{m-1}-2^{\lfloor m/2\rfloor}$, then the nonlinearity of $f(x,y)$ constructed by Theorem \ref{the bent A 4.1} exceeds $2^{n+m-1}-2^{\lfloor (n+m)/2\rfloor}$.
    If $m$ is even, $k>m/2-2$ and $g_1,g_2,g_3$ and $g_1\oplus g_2\oplus g_3$ are $m$-variable $k$-resilient functions achieving Sarkar et al's bound, then $N_f=2^{n+m-1}- 2^{n/2-1+k+1}$;
     If $m$ is even, $k\leq m/2-2$ and $g_1,g_2,g_3$ and $g_1\oplus g_2\oplus g_3$ are $m$-variable $k$-resilient functions achieving Sarkar et al's bound (their nonlinearity equal $2^{m-1}-2^{m/2-1}-2^{k+1}$), then $N_f=2^{n+m-1}- 2^{(n+m)/2-1}-2^{n/2+k+1}$, further, when $n=6$,  we can obtain a $(m+6)$-variable $k$-resilient function with nonlinearity $2^{6+m-1}- 2^{(6+m)/2-1}-2^{k+4}$.
       However, $f$  does not
    achieve Sarkar et al.'s  bound with equality, in general. }\end{remark}

\textbf{Examples of application}.  In \cite{Carlet AAECC,Carlet-Yucas2005} is given
an example of functions $f_1,f_2,f_3$ satisfying a condition which is the same as that needed in
Theorem \ref{the bent A 4.1}. Let $\vartheta(x)$ and $\theta(x)$ be
$n$-variable bent functions. Assume that there exists a vector $a$
such that $D_a\vartheta=D_a\theta$, where $D_a\vartheta(x)=\vartheta(x)\oplus \vartheta(x\oplus a)$
is the so-called derivative of $\vartheta$ at $a$.  We can take
$f_1(x)=\vartheta(x), f_2(x)=\vartheta(x\oplus a), f_3(x)= \theta(x)$, the
hypothesis of Theorem \ref{the bent A 4.1} is satisfied:
$\nu_1(x)=D_a\vartheta(x)\oplus \theta(x)=D_a\theta(x)\oplus
\theta(x)=\theta(x\oplus a)$ is bent and we have
$\widetilde{\nu_1}(x)=\widetilde{\theta}(x)\oplus a\cdot
x=(\widetilde{f_1}\oplus \widetilde{f_2}\oplus
    \widetilde{f_3})(x)$.

For example, let $x=(x',x'')\in {\Bbb F}_2^n, x',x''\in {\Bbb F}_2^{n/2}$. Let $\phi$ be a
  permutation on ${\Bbb F}_2^{n/2}$ and $\rho_1, \rho_2$ be  two arbitrary  $n/2$-variable Boolean
  functions. Let us define the M-M bent
  functions
  $\vartheta(x)=x'\cdot \phi(x'')\oplus \rho_1(x'')$,
  $\theta(x)=x'\cdot \phi(x'')\oplus
  \rho_2(x'')$. Let $a'$ be any nonzero element
   of ${\Bbb F}_2^{n/2}$ and $a=(a',0,\ldots,0) \in {\Bbb F}_2^n$.
  Thus, we have  $D_a\theta=D_a\vartheta$, that is, functions
  $f_1(x)=\vartheta(x), f_2(x)=\vartheta(x\oplus a), f_3(x)= \theta(x)$ satisfy
  the condition
  of Theorem \ref{the bent A 4.1}.

 \begin{remark}\label{rm 4.1}
{\em  According to Lemma \ref{lem increase}, we know that
  $W_f(\alpha,\beta)=W_{g_1}(\beta) W_{f_1}(\alpha)$,
   or $ W_{g_2}(\beta) W_{f_1}(\alpha)$, or $ W_{g_3}(\beta) W_{f_1}(\alpha)$, or $W_{g_1\oplus g_2\oplus g_3}(\beta) W_{f_1}(\alpha)$.
    Thus, from Theorem \ref{the bent A 4.1}, an   $(n+r)$th-order plateaued function in $n+m$ variables
    can be obtained if $g_1$, $g_2$ , $g_3$  and
  $g_1\oplus g_2\oplus g_3$ are $r$th-order plateaued functions.}
\end{remark}
Another consequence of Lemma \ref{lem increase} is the following secondary construction:
\begin{proposition}\label{cor4.1}
Let $n~ (>6)$  be an even positive integer. Let  $m$  and $k$ be two integers such that $k<m-1$. Let $f_1(x),
   f_2(x)$ and $f_3(x)$ be bent functions in $n$ variables such that  $\nu_1=f_1\oplus f_2\oplus f_3$ is also a bent function
   and
    $\widetilde{\nu_1}=\widetilde{f_1}\oplus \widetilde{f_2}\oplus
    \widetilde{f_3}$.
    Let $p(y)$ and $q(y)$  be two $k$-resilient functions in $m$ variables.\\
    If $W_{f_1}({\bf{0}})=W_{f_2}({\bf{0}})=W_{f_3}({\bf{0}})$ or $W_{f_1}({\bf{0}})\neq W_{f_2}({\bf{0}})=W_{f_3}({\bf{0}})$, where ${\bf{0}}=(0,0\ldots,0)\in {\Bbb F}_2^n$, then we set
    $g_1(y)=p(y)$, $g_2(y)=q(y)$ and $g_3(y)=q(y)\oplus y_i$;\\
    If $W_{f_1}({\bf{0}})=W_{f_2}({\bf{0}})\neq W_{f_3}({\bf{0}})$ or $W_{f_1}({\bf{0}})=W_{f_3}({\bf{0}})\neq W_{f_2}({\bf{0}})$, then we set
    $g_1(y)=p(y)\oplus y_i$, $g_2(y)=q(y)\oplus y_i$ and $g_3(y)=q(y)$,
     where $i\in\{1,2,\ldots,m\}$.\\
  Then,  $f(x,y)$, defined as in Theorem \ref{the bent A 4.1}, is a $k$-resilient function in $n+m$ variables with nonlinearity:

    \begin{equation}\label{equ bent resilient}
      \begin{array}{c}
         N_{f}\geq 2^{n+m-1}-
         2^{n/2-1}\times\max\left\{\max\limits_{\beta\in {\Bbb F}_2^m}\{|W_{p}(\beta)|\},\right.
         \left. \max\limits_{\beta\in {\Bbb F}_2^m}\{|W_{q}(\beta)|\}\right\},
      \end{array}
    \end{equation}
   the equality holds if and only if the equality $f_1=f_2=f_3$ does not hold.
\end{proposition}

\begin{proof}
  Since  $p(y)$ (resp. $q(y)$) is a $k$-resilient $m$-variable function, the resiliency order of $p(y)\oplus y_i$ (resp. $q(y)\oplus y_i$) is at least $k-1$, that is, $W_{p(y)\oplus y_i}(\beta)=0$ (resp. $W_{q(y)\oplus y_i}(\beta)=0$) for any $wt(\beta)\leq k-1$.

  From Theorem \ref{the bent A 4.1}, the function $f$ is at least $(k-1)$-resilient. Now, we prove $f$ is a $k$-resilient function in $n+m$ variables.

   When $W_{f_1}({\bf{0}})=W_{f_2}({\bf{0}})=W_{f_3}({\bf{0}})$ or $W_{f_1}({\bf{0}})\neq W_{f_2}({\bf{0}})=W_{f_3}({\bf{0}})$,  we set
    $g_1(y)=p(y)$, $g_2(y)=q(y)$ and $g_3(y)=q(y)\oplus y_i$.
    Thus, $g_1$ and $g_2$ are $k$-resilient functions, $g_3$ (resp. $g_1\oplus g_2\oplus g_3$)
  is  at least $(k-1)$-resilient.  Let $(\alpha,\beta)\in {\Bbb F}_2^{n+m}$ and $wt(\alpha,\beta)= k$. There are two different cases to consider.
  \begin{enumerate}
    \item If $wt(\alpha)\geq 1$, then $wt(\beta)\leq k-1$.  Moreover, we know that $W_{g_1\oplus g_2\oplus g_3}(\beta)$ $=0$ and $W_{g_3}(\beta)=0$. Certainly, $W_{g_1}(\beta)=0$ and $W_{g_2}(\beta)=0$.  From Relation (\ref{Walsh-th2}), $W_f(\alpha,\beta)=0$.
    \item If $wt(\alpha)=0$, i.e., $\alpha={\bf{0}}$, then $wt(\beta)=k$. We know  $g_1$ and $g_2$ are $k$-resilient functions, i.e.,  $W_{g_1}(\beta)=0$ and $W_{g_2}(\beta)=0$. According to Lemma \ref{lem increase},
       we know that $W_f(\alpha,\beta)=W_{g_1}(\beta) W_{f_1}(\alpha)$ (resp. $W_f(\alpha,\beta)=W_{ g_2}(\beta) W_{f_1}(\alpha))$ if
        $W_{f_1}(\alpha)=W_{f_2}(\alpha)=W_{f_3}(\alpha)$ (resp. $W_{f_1}(\alpha)\neq W_{f_2}(\alpha)=W_{f_3}(\alpha)$).  Thus, we have that $W_f(\alpha,\beta)=0$.
  \end{enumerate}

   When  $W_{f_1}({\bf{0}})=W_{f_2}({\bf{0}})\neq W_{f_3}({\bf{0}})$
   or $W_{f_1}({\bf{0}})=W_{f_3}({\bf{0}})\neq W_{f_2}({\bf{0}})$,  we set
   $g_1(y)=p(y)\oplus y_i$, $g_2(y)=q(y)\oplus y_i$ and $g_3(y)=q(y)$.
   We can prove $W_f(\alpha,\beta)=0$ for $wt(\alpha,\beta)= k$ by using
    the same method as above.

  Relation (\ref{equ bent resilient}) is then straightforward.
   From Lemma \ref{lem increase}, the equality of
    Relation (\ref{equ bent resilient}) holds if and only if
    the equality $f_1=f_2=f_3$ does not hold.
\end{proof}

\begin{remark}
{\em
If $N_{p}=N_{q}$, then $ N_{f}= 2^{n+m-1}-
         2^{n/2-1}\times\max\limits_{\beta\in {\Bbb F}_2^m}\{|W_{p}(\beta)|\}$.
         If we choose  $p(y),q(y)$ from PW functions ( Patterson and Wiedemann in \cite{Patterson1983} proposed 15-variable Boolean functions with
  nonlinearity $2^{14}-2^{7}+2^4+2^2$, which are called PW functions),
        then an $(n+15)$-variable function with nonlinearity $ 2^{n+15-1}-
         2^{n/2+7-1}+2^{n/2+4-1}+2^{n/2+2-1}$ can be obtained by Proposition \ref{cor4.1}. The nonlinearity of functions constructed by this way is the best known.
         In addition,
 if we apply direct sum (resp. indirect sum) using as initial functions $p(y)$ and $f_i(x)$ (resp. $p(y), q(y)$, $f_i(x)$ and $f_j(x)$), where $i,j=1,2,3, i\neq j$, then the nonlinearity of
functions constructed this way equals
 $ 2^{n+m-1}-
         2^{n/2-1}\times\max\limits_{\beta\in {\Bbb F}_2^m}\{|W_{p}(\beta)|\}$  as well.
         If we do not consider the resilience of the constructed function $f(x,y)$, then we can set $g_1(y)=p(y), g_2(y)=q(y)$ and $g_3(y)=q(y)\oplus l(y)$, where $l(y)\in A_m$.

  In \cite{Fu2011},  Fu et al.  proposed   a method for constructing
    $k$-resilient functions in odd numbers of variables. For odd $n\geq 35, k=1$ (resp. $n\geq 39, k=2$),
   a large class of  $k$-resilient $n$-variable functions, whose nonlinearity is  the best known,  can be constructed by the method.
   From their construction \cite[Construction]{Fu2011},  we found that the direct sum
functions were chosen  initial functions. Here, if we substitute  the functions constructed by Proposition \ref{cor4.1}  for the direct sum functions,  then
    many resilient functions on odd number of variables whose nonlinearities equal those of the functions presented by Fu et al. in \cite{Fu2011} can be obtained. }
\end{remark}

\begin{example}\label{example1}
 Several constructions of 8-variable 1-resilient functions with nonlinearity $116$ were presented in \cite{Clark2002,S. Maitra2002,S. Maity2002,S. Maity2004}.  By using two different
  $1$-resilient $8$-variable functions and three $6$-variable bent functions $f_1,f_2,f_3$
  (which satisfy $f_1\oplus f_2\oplus f_3$ being also bent and
   $\widetilde{f_1\oplus f_2\oplus f_3}=\widetilde{f_1}\oplus \widetilde{f_2}\oplus \widetilde{f_3}$),
  with Proposition \ref{cor4.1}, we can obtain $14$-variable 1-resilient
   functions with nonlinearity $2^{13}-2^{6}-2^{5}=8096$.
 The functions $(14,1,-,8096)$ earlier known could only be obtained by direct sum and indirect sum.
 \end{example}


  Clearly, the functions constructed by  Proposition \ref{cor4.1} are different from those  constructed by direct sum. In Table \ref{tab:1}, we describe  the difference between
the functions constructed by Proposition \ref{cor4.1} and the functions  constructed by indirect sum.

  \begin{table*}[!t]
\renewcommand{\arraystretch}{1.3}
\caption{Forms of Functions Constructed by Indirect Sum and Proposition \ref{cor4.1}}
\label{tab:1}
\centering
\begin{tabular}{|c|c|c|} \hline
Initial Functions &Indirect sum  &Proposition \ref{cor4.1}\\\hline

$f_1\neq f_2, f_2\neq f_3$, & $f_1(x)\oplus g_1(y)\oplus $  &  $f_1(x)\oplus g_1(y)\oplus$\\

$f_1\neq f_3$, $g_1\neq g_2$, &  $(f_1\oplus f_2)(x)(g_1\oplus g_2)(y)$   &  $(f_1\oplus f_2)(x)(g_1\oplus g_2)(y)$  \\

  $g_3=g_2\oplus y_i$  &    &  $\oplus  y_i(f_2\oplus f_3)(x)$\\ \hline

$f_1\neq f_2, f_2\neq f_3$, &$f_1(x)\oplus g_1(y)\oplus $&  $f_1(x)\oplus g_1(y)\oplus $\\
 $f_1= f_3$,$g_1\neq g_2$,&$(f_1\oplus f_2)(x)(g_1\oplus g_2)(y)$ &  $(f_1\oplus f_2)(x)(g_1\oplus g_2)(y)$ \\
   $g_3=g_2\oplus y_i$  & &$\oplus y_i(f_1\oplus f_2)(x)$ \\ \hline

\end{tabular}
\end{table*}

\section{Conclusion}\label{5}
 Bent functions and resilient functions with high nonlinearity
are actively studied for their numerous applications in
cryptography, coding theory, and other fields.

In this paper, we focused  on the constructions of both bent functions and highly nonlinear Boolean functions.   We first presented a novel secondary construction of bent functions. By using this method, we could  deduce several
concrete constructions of bent functions from known bent functions.  In addition,
we presented a generalization of the indirect sum construction for constructing resilient functions with high nonlinearity.


\section{Acknowledgment}
{This work was supported in part by National Science Foundation of
China (60833008, 61173152), and in part
Science and Technology on Communication Security Laboratory
(9140C110201110C1102).}

\end{document}